\documentclass[a4paper,fleqn]{cas-dc-arxiv}

\usepackage[numbers]{natbib}
\usepackage{lineno,hyperref}

\usepackage{epsfig}

\usepackage{amssymb}
\usepackage{amsthm}

\usepackage{subfigure}
\usepackage{graphics}
\usepackage{multicol}
\usepackage{graphicx}

\modulolinenumbers[5]

\def\tsc#1{\csdef{#1}{\textsc{\lowercase{#1}}\xspace}}
\tsc{WGM}
\tsc{QE}
\tsc{EP}
\tsc{PMS}
\tsc{BEC}
\tsc{DE}

\begin{document}
\let\WriteBookmarks\relax
\def\floatpagepagefraction{1}
\def\textpagefraction{.001}
\shorttitle{Unknotting of ferrogranular networks under field}
\shortauthors{P.A. S\'anchez et~al.}

\title [mode = title]{Unknotting of quasi-two-dimensional ferrogranular networks by in-plane homogeneous magnetic fields}                      



\author[1,2]{Pedro A. S\'anchez}[orcid=0000-0003-0841-6820]
\cormark[1]
\ead{r.p.sanchez@urfu.ru}

\credit{TBD} 

\address[1]{Ural Federal University, 51 Lenin av., Ekaterinburg, 620000, Russian Federation.}
\address[2]{Institute of Ion Beam Physics and Materials Research, Helmholtz-Zentrum Dresden-Rossendorf e.V., D-01314 Dresden, Germany.}

\author[3]{Justus Miller}
\ead{justus.miller@uni-bayreuth.de}

\credit{TBD} 

\address[3]{Experimentalphysik 5, University of Bayreuth, 95440 Bayreuth, Germany.}

\author[1,4]{Sofia S. Kantorovich}[orcid=0000-0001-5700-7009]
\ead{sofia.kantorovich@univie.ac.at}

\credit{TBD} 

\address[4]{Computational Physics, University of Vienna, 1090 Vienna, Austria.}

\author[3]{Reinhard Richter}[orcid=0000-0002-6380-5477]
\ead{reinhard.richter@uni-bayreuth.de}

\credit{TBD} 

\cortext[cor1]{Corresponding author}

\begin{abstract}
Our ongoing research addresses, by means of experiments and computer simulations, the aggregation process that takes place in a shaken granular mixture of glass and magnetized steel beads when the shaking amplitude is suddenly decreased. After this quenching, the steel beads form a transient network that coarsens in time into compact clusters, following a viscoelastic phase separation. Here we focus on the quasi-two-dimensional case, analyzing in computer simulation the effects of a magnetic field parallel to the system plane. Our results evidence that the field drastically changes the structure of the forming network: chains and elongated clusters parallel to the field are favored whereas perpendicular connecting structures tend to be supressed, leading to the unknotting of the networks observed at zero field. Importantly, we found that moderate field strengths lead to the formation of larger clusters at intermediate time intervals than in the case of weak and strong fields. Moreover, the latter tend to limit the overall growth of the clusters at longer time scales. These results may be relevant in different systems governed by similar magnetically driven aggregation processes as, for example, in the formation of iron-rich planetesimals in protoplanetary discs or for magnetic separation systems.
\end{abstract}


\maketitle

\sloppy

\section{Introduction}
Field induced aggregation of magnetizable particles is a relevant process in different systems and length scales. Whereas large aggregation of magnetic nanoparticles in ferrofluids ($10^{-8}$m) is usually undesirable \cite{rosensweig1985}, it plays a decisive role in magnetorheological fluids ($10^{-5}$m) \cite{furst2000dynamics,rinaldi2005b} as well as in purification techniques for water \cite{ambashta2010water} or air \cite{li2007aggregation}. On a macroscopic level  ($ 10^{-3}$m) it is important for ore separation \cite{iranmanesh2017magnetic} and may even be an important mechanism in the formation of iron-rich planetesimals, precursors of rocky planets in protoplanetary discs \cite{nuth1994magnetically,dominik2002magnetic,nubold2003magnetic,2018-kruss-aphj}.

\begin{figure}[]
\centering
  \includegraphics[width=\columnwidth]{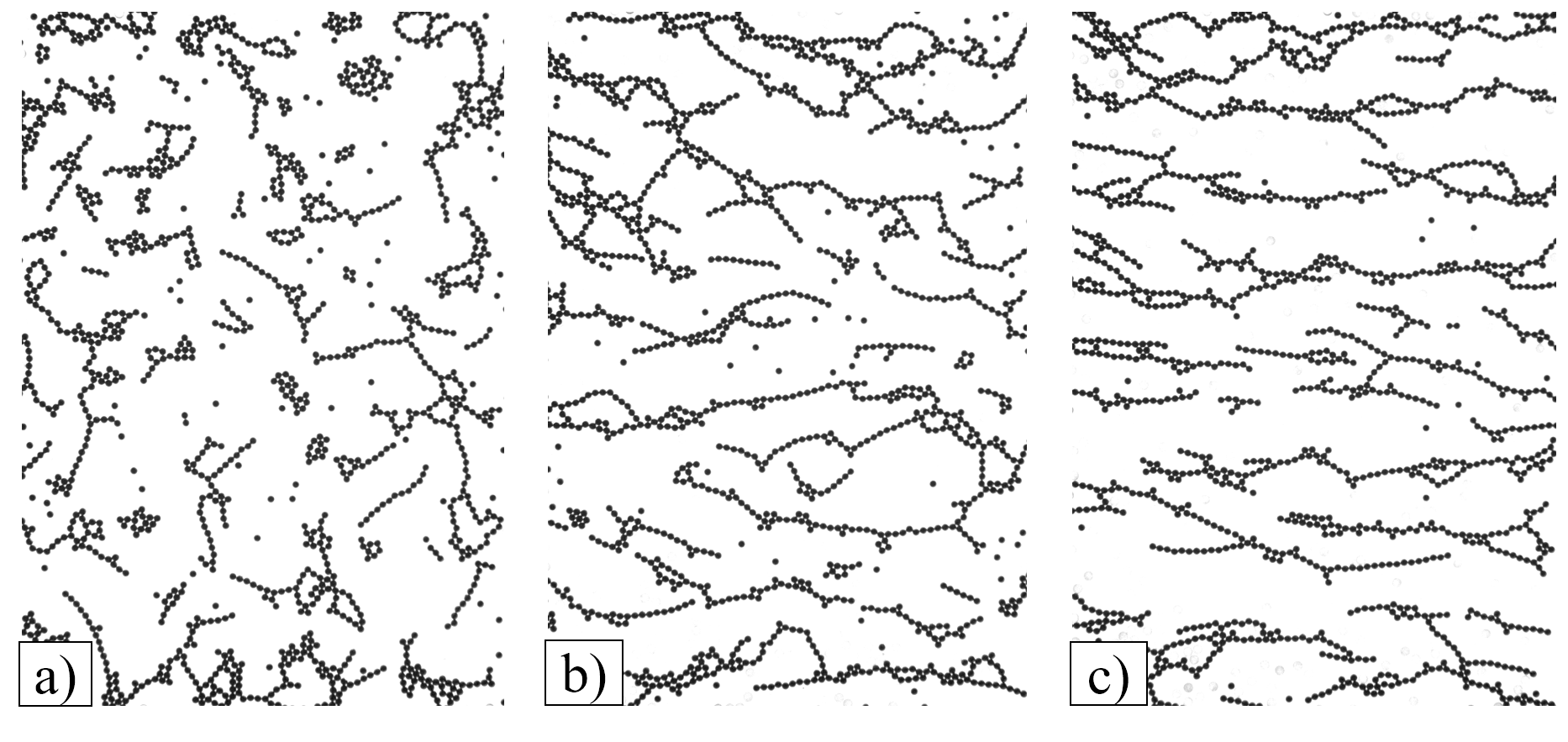}
  \caption{Network formation of steel beads 4\,s after the quench of the shaker amplitude from $\Gamma$=3.0g to 1.8g without (a) and with an externally applied field of $B=0.2$\,mT (b)  $0.7$\,mT (c). Note that the glass beads have a low contrast and therefore can not be unveiled here.}
  \label{fig:exp_snapshot}
\end{figure}

The influence of an external field $\vec{H}$ on the aggregation of particles that are already magnetized is still not fully understood. An interesting model system suitable for the fundamental study of such phenomena is the granular mixture of glass and magnetized steel beads. Recently, we have shown in experiment and simulation \cite{2018-koegel} that, even without an external field, a quasi-two-dimensional shaken system of such mixture tends to demix when the shaking amplitude is not too high, as the steel beads form aggregates. This corresponds to a ferrogranular coarsening dynamics that evolves in three stages. In an initial phase (i) chain-like small aggregates are forming, leading to the emergence of a network with a significant fraction of chains, as shown in Fig.\,\ref{fig:exp_snapshot}(a); such loose network evolves by forming regions of close packed arrangements, whereas long chains and other thin structures tend to break (ii); finally, only large compact clusters of magnetized beads with rounded profiles tend to remain (iii). This scenario resembles the initial (i), elastic (ii) and hydrodynamic (iii) phases of a viscoelastic phase separation (VPS). VPS was first uncovered by Hajime Tanaka \cite{tanaka2000, tanaka2005} in suspensions of polymers in a much less viscous solvent, like water. The driving forces for the VPS are the differences in time scales of the constituents. The polymers, or in the present case, the magnetized spheres forming aggregates, have a slower dynamics than the polymer solvent or the non-aggregating glass beads.

The aggregation forces between steel beads have two contributions. One is the interaction between their permanent magnetic moments, well represented by the dipole-dipole pair potential. The second stems from the fact that these beads are also magnetically susceptible. We may, therefore, address them as susceptible dipolar hard spheres (SDHS). These SDHS respond sensitively to the field originating from the neighbouring beads - see Fig.\,\ref{fig:exp_snapshot}a - as well as to an externally applied field, as shown in Figs.~\ref{fig:exp_snapshot}b and \ref{fig:exp_snapshot}c.

In the following, we investigate the impact of an in-plane homogeneous magnetic field $\vec{H}$, generated by a Helmholtz-pair of coils \cite{miller2019messungen} on the formation of transient networks and their coarsening dynamics. In this work we put our emphasis on simulation predictions, which are only qualitatively compared to experimental data.  

\section{Modeling approach}
In order to characterize the experiment, described in detail in Ref.\,\cite{2018-koegel}, we introduced a phenomenological numerical model that mimics the coarsening process of aggregates of SDHS by means of Langevin dynamics simulations. In this approach, glass and steel beads are represented as soft-core spheres whose movement is constrained to a plane with lateral periodic boundaries. The effects of the mechanical shaker are mimicked by the thermal noise of the Langevin thermostat, so that the experimental amplitude of the shaker corresponds to the system temperature in simulations and the quenching is performed by suddenly decreasing its value. The magnetic interactions between the steel beads are simulated using a minimal approach that combines a point dipole placed in the center of each magnetic sphere, $\vec \mu$, and an isotropic Lennard-Jones attraction of strength $\epsilon_a$ and radius given by the size of the particles. In this way, each pair of magnetic beads experiences the conventional anisotropic dipole-dipole interaction plus a central attraction. The parameters of this model and the system of reduced units in which its results are expressed were fitted in \cite{2018-koegel} by comparing the short time evolution of the mean coordination number, or degree, of the networks of magnetic particles formed immediately after quenching.

Here, we use the same model and fitted reduced parameters as in \cite{2018-koegel} in order to study the effects of introducing a magnetic field, $\vec H$, during the coarsening process. Briefly, such reduced values are, respectively, $\sigma_\mathrm{m}=3$ and $\sigma_\mathrm{g}=4$ for the diameters of the magnetic and non magnetic particles, $\phi_\mathrm{m}=0.18$ and $\phi_\mathrm{g}=0.15$ for their area fractions, $T_1=5.0$, and $T_2=0.5$ for the temperatures before and after quenching, $r_\mathrm{cut}=1.2 \sigma_\mathrm{m}=3.6$ the cutoff distance to consider two magnetic particles to be in close contact, and $\lambda = \mu^2 / T_2 \sigma_\mathrm{m}^3=5$ is the conventional measure of the strength of the dipole-dipole interaction after quenching.

The application of the external field introduces a new interaction, represented by the Zeeman potential acting on each magnetic particle $i$ according to its corresponding dipole, $\vec \mu_i$, via
\begin{equation}
    U_Z= - \vec \mu_i \cdot \vec H.
    \label{eq:U.H}
\end{equation}
This also introduces new free parameters in the model that should be fitted in order to allow any quantitative comparison between experiments and simulations. Here, however, we aim only at a qualitative analysis, sampling an arbitrary range of reduced values of $H=\| \vec H \|$. Proper quantitative model fitting will be performed in a future work.

Complete details of the simulation protocol can be found in Ref.\,\cite{2018-koegel}. Simulations were performed with the {ESPResSo} 3.3.1 simulation package \cite{2013-arnold}.

\section{Results and discussion}

\begin{figure}[]
\centering
  \includegraphics[width=8.2cm]{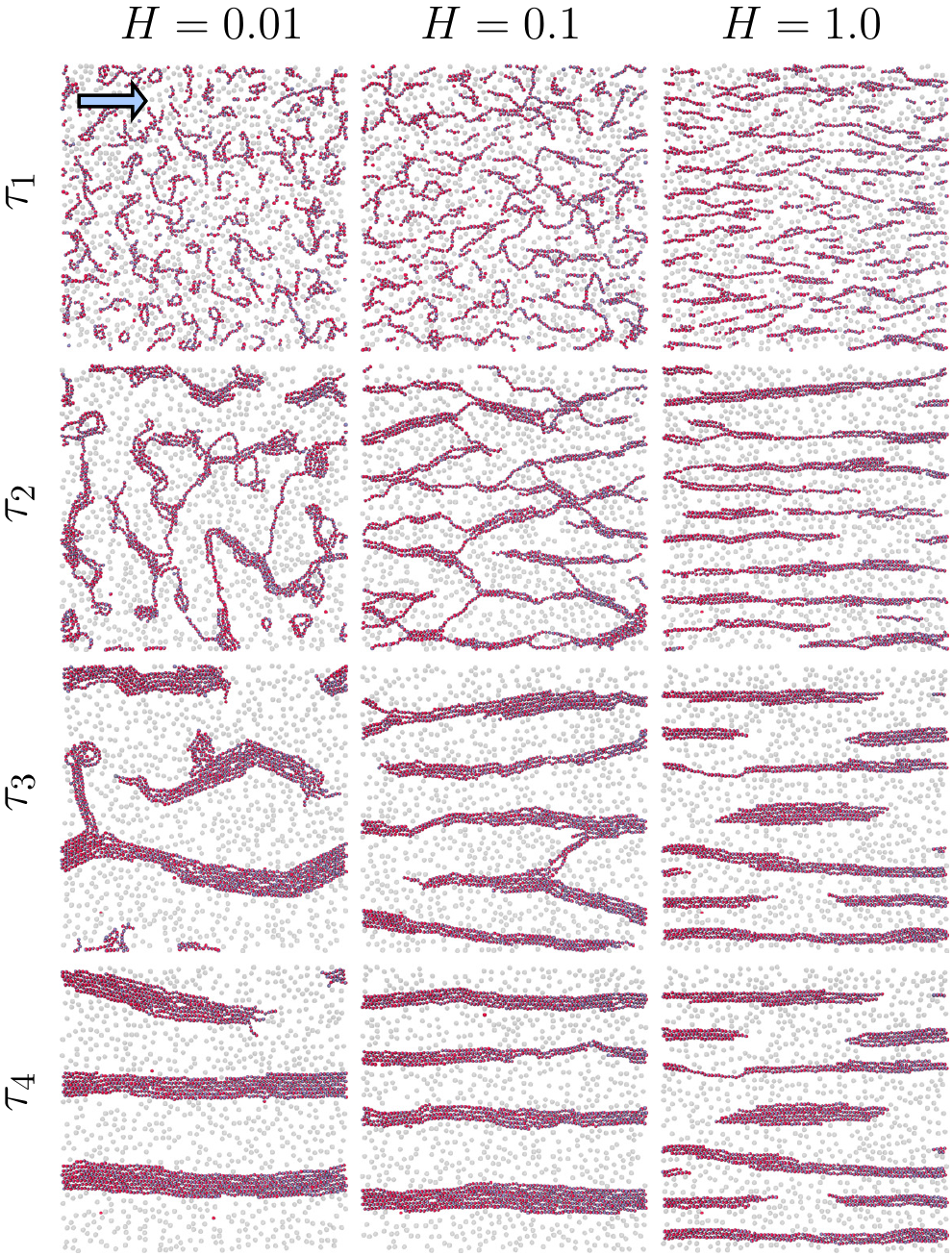}
  \caption{Simulation snapshots for increasing magnetic fields (left to right) and increasing time (top to bottom). Magnetic particles are dark colored, non magnetic are light grey. Arrow in the top left panel indicates the orientation of the field in all cases. Simulation time values: $\tau_1=34$ (estimated initial regime), $\tau_2=1140$ (est. elastic regime), $\tau_3=14000$ (est. early hydrodynamic regime), $\tau_4=64000$ (est. late hydrodynamic regime).}
  \label{fig:simsnaps}
\end{figure}

Figure~\ref{fig:simsnaps} shows the time evolution of the networks (top to bottom) for three selected values of an in-plane applied magnetic field. The influence of a weak field $H=0.01$ (left column) can be hardly appreciated at early stages ($\tau_1$ and $\tau_2$), whereas the compact clusters at $\tau_3$ already evidence their alignement with $\vec{H}$, and at $\tau_4$ no connecting structures perpendicular to $\vec{H}$ persist. This ``unknotting'' effect of $\vec{H}$ becomes more prominent for the intermediate field $H=0.1$ (middle column), where already at $\tau_2$ the network is clearly oriented. For a strong field $H=1.0$, the formation of a two-dimensional network is hindered already at $\tau_1$ and only well aligned clusters exist. This arises the question, whether for large $H$ an elastic regime, characterized by a coarsening two-dimensional network, could exist at all.

\begin{figure}[]
\centering
  \includegraphics[width=8.5cm]{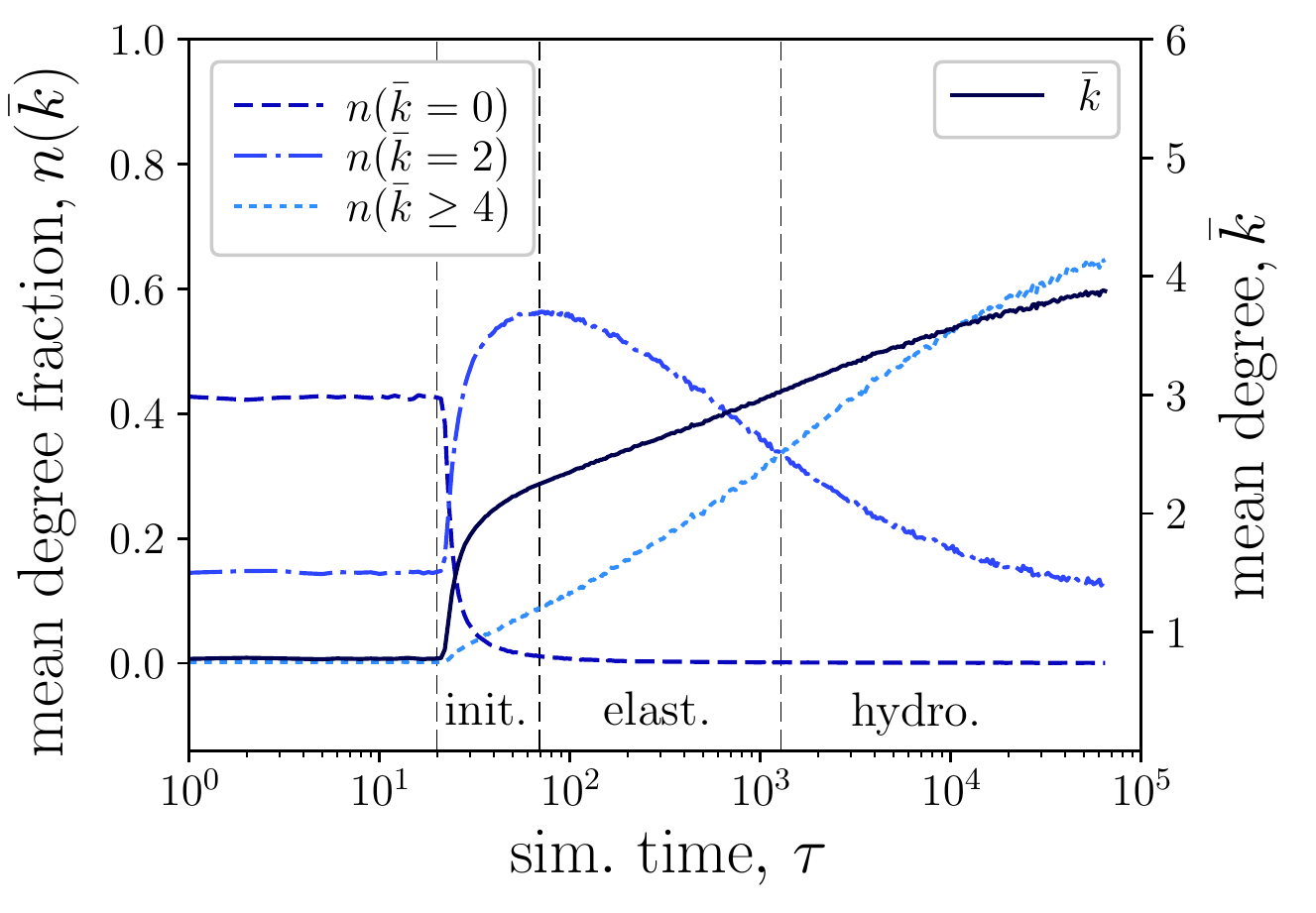}
  \caption{Identification of the boundaries (vertical dashed lines) of the initial (init.), elastic (elast.) and hydrodynamic (hydro.) regimes of the viscoelastic transition from the time evolution of fractions of magnetic particles with different partial mean degrees, $n\left ( \bar k \right )$, corresponding to a selected value of applied field, $H=0.1$. The vertical dashed lines at $\tau_{\mathrm{in}}=20$ (quenching time), $\tau_{\mathrm{el}}=69$ (maximum of $n(\bar k)=2$) and $\tau_{\mathrm{hy}}=1280$ (crossing between $n(\bar k=2)$ and $n(\bar k\ge 4)$), mark the onset of the three regimes. 
  The curve for the mean degree, $\bar k$, is also shown as reference. }
  \label{fig:meandegreefractions}
\end{figure}

\begin{figure}[]
\centering
  \includegraphics[width=7.6cm]{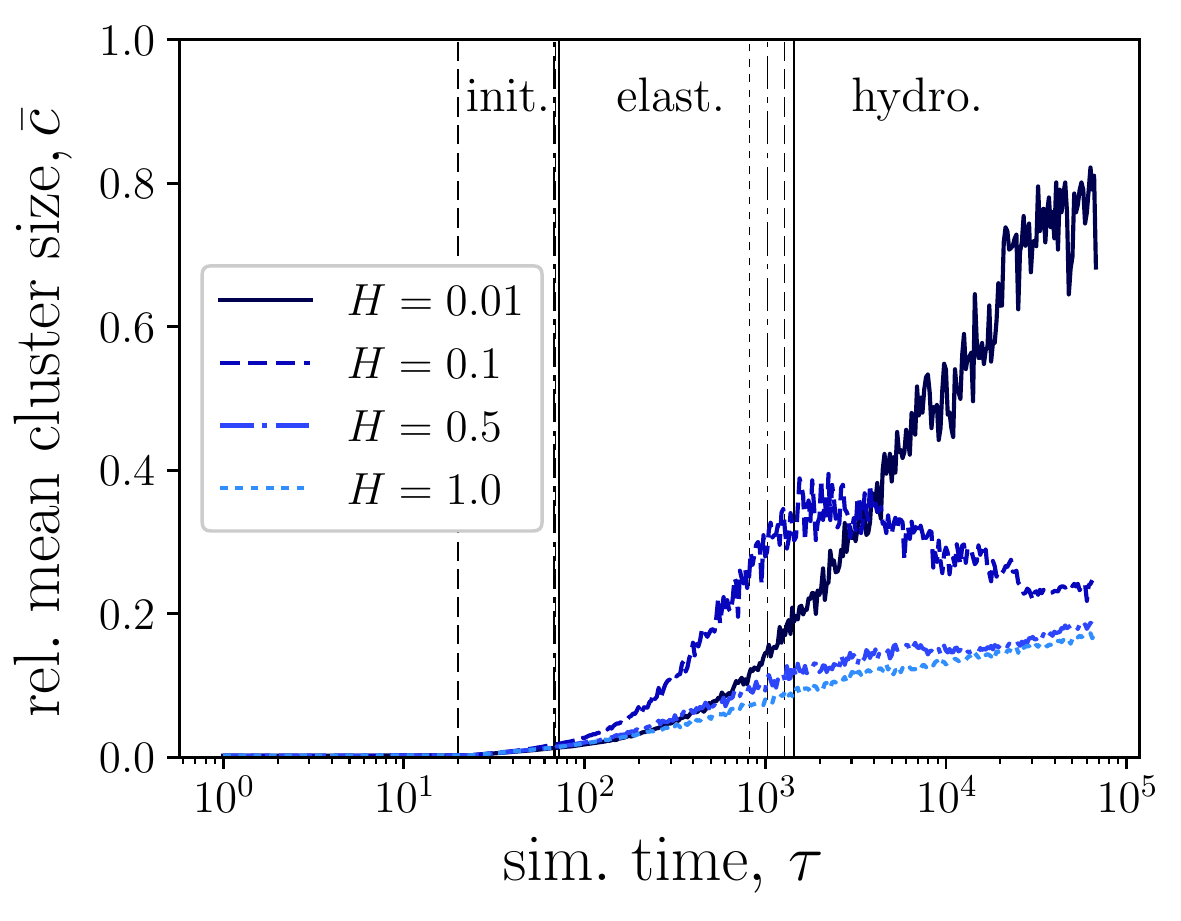}
  \caption{
  Evolution of the cluster size for four different values of the applied magnetic field. Vertical lines indicate the estimated borders between VPS regimes corresponding to each field.}
  \label{fig:clustersize}
\end{figure}

Here we introduce a criterium to roughly identify the VPS regimes in our system. By definition, the initial regime is starting immediately after the quenching takes place, $\tau_{\mathrm{in}}$. For the rest, we need to find proper characteristic parameters. As an example, Figure~\ref{fig:meandegreefractions} shows the evolution of different network parameters measured for $H=0.1$. The mean degree ($\bar k$, solid line) does not provide a clear indication of the onset of the elastic or hydrodynamic regimes. However, the fraction of magnetic particles with degree 2 ($n(\bar k=2)$, dashed-dotted line), is a good measure of the amount of chain-like structures in the system. In the initial regime, $n(\bar k=2)$ has a steep growth, whereas the fraction of non-clustered magnetic particles ($n(\bar k=0)$, dashed line) strongly decays. However, $n(\bar k=2)$ has to decay in the elastic regime. Therefore, we take the maximum of $n(\bar k=2)$, $\tau_{el}$, as the border between the initial and elastic regimes. Finally, compact clusters are mainly formed by particles with degree not lower than 4. Note that the corresponding fraction $n(\bar k\ge4)$ (dotted line) grows monotonically. Therefore, we can consider that the crossover to the hydrodynamic regime, $\tau_{\mathrm{hy}}$, takes place when compact clusters start to dominate, signaled by the crossing between $n(\bar k=2)$ and $n(\bar k=2)$. The effect of field strengths $H \in[0.01,1]$ on these regime borders (vertical lines in Fig.~\ref{fig:clustersize}) is the following: first, the elastic regime tends to narrow with field strength, as $\tau_\mathrm{hy}$ decreases by 43\% as $H$ increases; surprisingly, $\tau_\mathrm{el}$ is instead hardly affected ($<5\%$), which is not in accordance with the visual impression of Fig.\ref{fig:simsnaps}. Obviously, these definitions are based on isotropic averages of scalar parameters, $n(\bar k)$, which can not take into account the different evolution in directions parallel and perpendicular to the field.

The impact of the field is more clearly observed in the evolution of the relative mean cluster size $\bar{c}$, as shown in Fig.~\ref{fig:clustersize}. In the initial regime, the growth of $\bar{c(\tau)}$ can hardly be discriminated for different values of $H$. In the elastic regime, the intermediate field $H=0.1$ (dashed line) produces the strongest growth, but it reaches a maximum around $\tau_\mathrm{hy}$ and subsequently decays. This is explained by this field being strong enough to prevent the formation of loops (as observed for weak or zero fields) but not enough to prevent the existence of connecting structures perpendicular to the field during the elastic regime (see $\tau_1$ to $\tau_3$ in Fig.~\ref{fig:simsnaps}). This leads to a high connectivity and large cluster sizes, that break down in the hydrodynamic regime ($\tau_4$ in Fig.~\ref{fig:simsnaps}). In this latter regime, largest clusters correspond to the weakest field, as the orientation is not so strongly constrained. Large fields, instead, strongly hinder not well aligned structures, making the growth of the clusters to become dynamically arrested.

\begin{figure}[]
\centering
  \includegraphics[width=7.5cm]{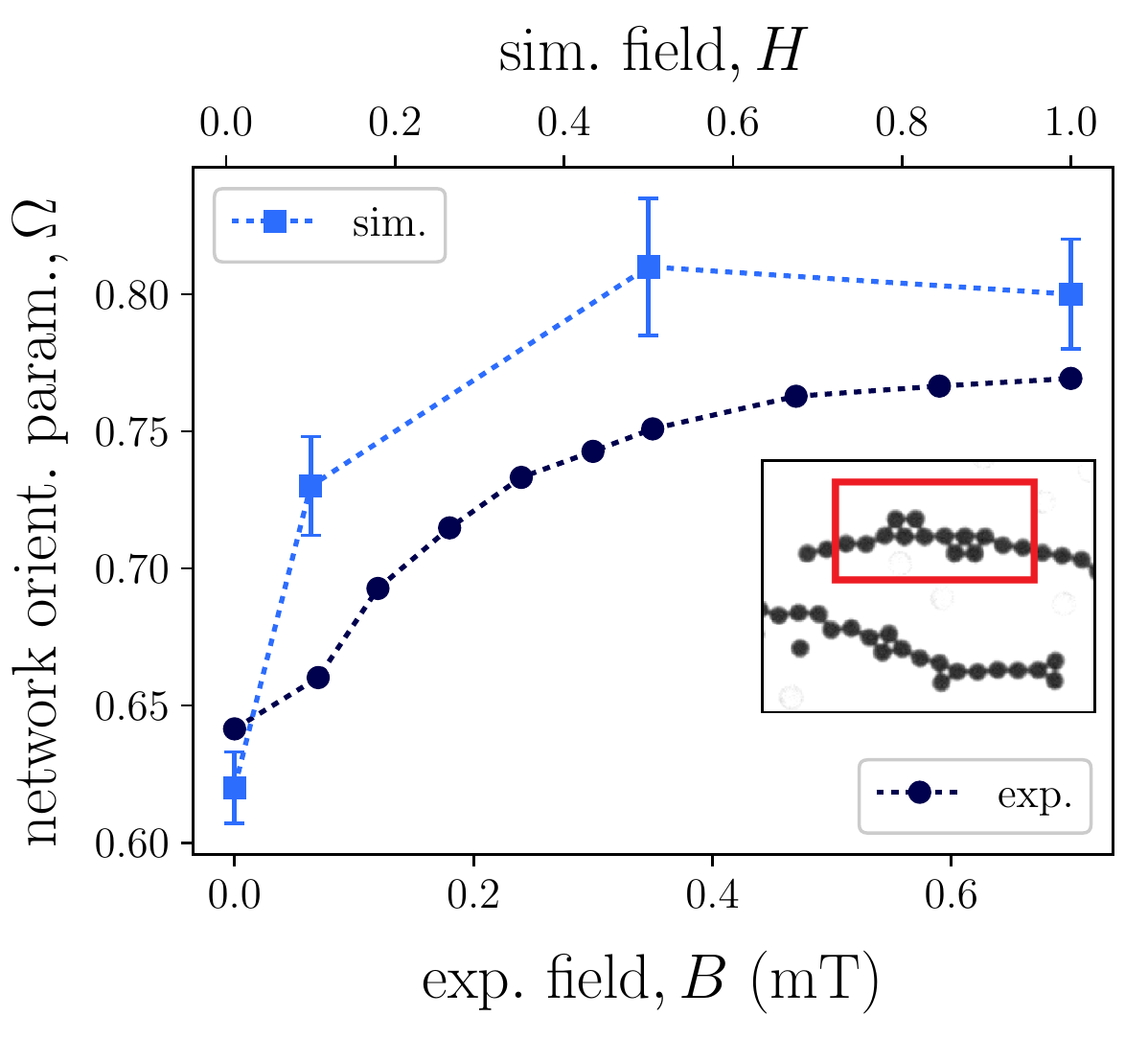}
  \caption{The network orientation parameter $\Omega$ (\ref{eq:omega}) vs the simulated field $H$ (top axis) and vs experimental applied induction $B$ (bottom axis). Simulation data (light squares) corresponds to $\tau=120$, approximately the estimated border between initial and elastic regimes. Experimental data (dark circles) was obtained from frames recorded 4\,s after the quenching, from $\Gamma=3.6$g to 1.8\,g. Dotted lines are a guide for the eye. The inset shows the detail of an experimental network recorded at $B=1.4$mT. The red frame underlines examples of close packed regions.
  }
  \label{fig:omega}
\end{figure}

In order to capture the field-induced symmetry breaking observed in experiments (Fig.\,\ref{fig:exp_snapshot}) and simulations (Fig.\,\ref{fig:simsnaps}), we measured the angles $\theta_i$ between the center-to-center vector of any pair of magnetized beads in close contact and the applied field. From such angles we define the orientational parameter
\begin{equation}
 \Omega = \frac{\sum_{\theta_i} \cos(\theta_i) n(\theta_i)}{\sum_{\theta_i} n(\theta_i)}.
 \label{eq:omega}
\end{equation}
$\Omega$, that is easily accesible from both, experimental and simulation data, becomes $2/\pi \approx 0.64$ for isotropic structures and approaches 1 for simple pearl necklace chains ($\bar{k}=2$) perfectly aligned with the field. Fig.~\ref{fig:omega} shows experimental and simulation results for $\Omega$ as a function of the applied field. Even the scale of the field in simulations is not fitted to the experimental values, both measurements show the same qualitative trend: $\Omega$ grows with field strength from nearly its isotropic value at zero field up to a saturation value. However, the latter does not correspond to the maximum expected value 1. This is due to the field slightly favoring the formation of compact regions with hexagonal order at early times (see inset in Fig.~\ref{fig:omega}).

\section{Summary and conclusions}
We have investigated the effects of an in-plane applied field on the coarsening dynamics of a shaken mixture of glass and magnetized steel beads. In both, experiments and computer simulations, the applied field is breaking the symmetry of the transient networks of aggregated magnetic particles that emerge after a quenching of the shaking amplitude, favoring the formation of elongated structures parallel to its orientation. For increasing $H$, this effect becomes more prominent at short time scales, as a network orientational parameter reveals in both approaches.The effects at intermediate and long time scales have been studied in computer simulations only. A non obvious result is that intermediate fields favor more strongly the growth of the aggregates at intermediate time scales. At long time scales the largest clusters are found for the weakest sampled field, whereas strong fields tend to hamper their growth.  

These findings may be of relevance for the cluster growth in planetesimals, precursors of planets. They do not support the conjecture that 
an external magnetic field monotonically favors the cluster growth \cite{2018-kruss-aphj}. However, our observations are still preliminary, as Eq.~(\ref{eq:U.H}) does not take into account a field dependent susceptibility of the magnetized spheres. A quantitative comparison with experimental results will be included in a forthcoming study. 

Moreover, we have analyzed how $H$ is shifting the onset for the initial, elastic and hydrodynamic regimes of the viscoelastic phase separation investigated previously \cite{2018-koegel}. As an order parameter we tested the fractions of magnetic particles with given degree. From these isotropically averaged parameters, only a significant dependence for the boundary between intermediate and long time regimes was found. In forthcoming investigations this approach must be replaced by a measure that can discriminate parallel and perpendicular directions with respect to the field.

\section*{Acknowledgements}
Research supported by the Russian Science Foundation Grant No.19-12-00209. Simulations were performed at the Vienna Scientific Cluster (VSC3). R.R. gratefully acknowledges I. Rehberg for supporting his attendance to ICMF19.

\end{document}